\documentclass[prl,twocolumn,showpacs,preprintnumbers]{revtex4}
\usepackage{graphicx}
\usepackage{dcolumn}

\newcommand{\beq}{\begin{equation}}
\newcommand{\eeq}{\end{equation}}
\newcommand{\beqa}{\begin{eqnarray}}
\newcommand{\eeqa}{\end{eqnarray}}

\begin{document}

\title{Evolution of the dynamical pairing across
the phase diagram of \\
a strongly correlated high-temperature superconductor}
\author{M. Civelli$^1$}
\affiliation{$^1$ Theory Group, Institut Laue Langevin, 6 rue
Jules Horowitz 38042 Grenoble Cedex, France}

\begin{abstract}
We study the dynamics of the Cooper pairing across the $T=0$ phase
diagram of the two-dimensional Hubbard Model, relevant for
high-temperature superconductors, using a cluster extension of
dynamical mean field theory. We find that the superconducting
pairing function evolves from an unconventional form in the
over-doped region into a more conventional boson-mediated
retarded form in the under-doped region of the phase diagram. The
boson, however, promotes the rise of a pseudo-gap in the electron
density of states rather than a superconducting gap as in the
standard theory of superconductivity. We discuss our results in
terms of Mott-related phenomena, and we show that they can be
observed in tunneling experiments.
\end{abstract}

\date{\today}
\maketitle

In order to understand high-temperature (high-Tc)
superconductivity one must understand the nature of the pairing
interaction forming the Cooper pairs. Pairing in standard
superconductors has been successfully described by the
Migdal-Eliashberg (ME) formalism\cite{migdal58-eliashberg60}, an
extension of the famous Bardeen-Cooper-Schrieffer theory (BCS) of
superconductivity\cite{bcs}. The essential ingredient of the ME
description is a retarded boson-mediated interaction between
electrons\cite{scalapino69}. It has been argued, however, that
high-Tc superconductivity is the product of a strongly correlated
mechanism\cite{anderson87}, which gives rise to an instantaneous
electron pairing\cite{anderson07}, radically different from the
ME description. Recent experimental\cite{yazdani08,heumen08,maksimov08}
and theoretical\cite{maier08,markiewicz08,haule07} studies on high-Tc
superconductors have focused on this problem, but a number of
questions remain unanswered. In this paper we enter the debate,
examining the pairing interaction as one moves from the over-doped (OD)
region to the under-doped (UD) region of the phase diagram of
the two-dimensional Hubbard Model (a minimal model containing the
physics of high-Tc superconductors\cite{anderson87}).
We look beyond the BCS and ME theories, using the cellular dynamical
mean field theory (CDMFT\cite{kotliar06,cluster05-06}). The CDMFT can
address the full frequency-dependence of the pairing interaction and the
short-ranged spatial correlation.

It is well established\cite{general04} by both
experiment\cite{damascelli03-campuzano04} and
theory\cite{anderson87} (including CDMFT
studies\cite{haule07,marce08b}) that high-Tc superconductors have
many properties that are BCS-like in the OD region and anomalous
in the UD region. Little, however, is known about the dynamics of
the Cooper pairing, it being extremely difficult to identify its
contribution to physical observables\cite{scalapino69,yazdani08}.
We show that in the OD region the pairing function does not
display BCS or ME forms. Despite the presence of a boson-mediated
pairing contribution at low frequencies, other features are
relevant up to an energy scale $W^{\ast}$ of the order of the
bandwidth reduced by the strong interaction. In the UD region,
however, the pairing function acquires an ME form, even if at
higher energies (but within $W^{\ast}$) a pair-breaking
contribution appears. We connect these findings with the
emergence of a pseudo-gap in the electron spectra at optimal
doping (competing with the superconducting gap) and we interpret
this in terms of Mott-related phenomena\cite{marce08,marce08b}.
Finally we show that these properties can be observed in scanning
tunneling microscopy (STM) experiments. The two-dimensional
Hubbard Model Hamiltonian is:
\begin{equation}
\mathcal{H} = -\sum_{i,j,\sigma}t_{ij}\, c^{\dagger}_{i\sigma}
c_{j\sigma}  + U \sum_{i} n_{i\uparrow}n_{i\downarrow}-\mu \,
\sum_{i\sigma}\, n_{i\sigma} \label{hamiltonian}
\end{equation}
where $c_{i\sigma}$ destroys an electron with spin $\sigma$ on
site $i$ and $n_{i\sigma}= c^{\dagger}_{i\sigma} c_{i\sigma}$. We
only consider the nearest-neighbor $t$ and next-nearest-neighbor
$t^{\prime}= -0.3t$ hoppings; $\mu$ is the chemical potential. We
set the on-site Coulomb repulsion $U=12t$, greater than the
bandwidth $8t$. In CDMFT, $\mathcal{H}$ is mapped onto a more
easily handled impurity model (2$\times$2 cluster model in our
case) of interacting electrons, embedded in a non-interacting
bath of fermions, subjected to a self-consistency
condition\cite{kotliar06}. We solve the cluster-impurity problem
using the Lanczos algorithm\cite{krauth94}, which approximates the
non-interacting bath with an 8-level parameterization. This method
works at zero temperature\cite{note-method} and provides direct
access to the information on the real-frequency axis, unlike, for
example, the quantum-Monte-Carlo
method(QMC)\cite{hirsch86-werner06}.

In this study we consider a paramagnetic translationally invariant
superconducting state by constraining the CDMFT
equations. Even if other broken-symmetry phases compete for the
ground-state (at low doping, antiferromagnetism is expected to
take over, for example), this is a well-defined mean-field
procedure which allows us to access and study the physics
governing the superconductivity. We leave open the question of
which terms could be added to $\mathcal{H}$ to make this superconducting
mean-field solution into a real ground-state. The output of our
CDMFT calculation is the frequency-dependent normal
$G^{nor}_{ij}(\omega)= \ll c_{i\sigma} c^{\dagger}_{j\sigma} \gg$
and anomalous
$F_{ij}(\omega)= \ll c_{i\uparrow} c_{j\downarrow} \gg$
cluster Green's functions and their associated normal
$\Sigma^{nor}_{ij}(\omega)$ and anomalous $\Sigma^{ano}_{ij}(\omega)$
cluster self-energies ($i,j=1,...,4$ in the $2\times 2$
cluster\cite{venky05,marce08b}). We indicate the Fourier transform
of the time-ordered ground-state average with $\ll\ldots\gg$.
In this paper we focus on the nearest-neighbor anomalous component
$\Sigma^{a}_{}(\omega)$ (the only one numerically non-zero), which
directly expresses the pairing function. We think that it
captures the essential features of the superconducting pairing,
which has a d-wave momentum dependence $\sim (\cos k_x- \cos k_y)$
(dominated by the nearest-neighbor spatial
component\cite{poilblanc05}), as has been well established
experimentally\cite{damascelli03-campuzano04} and
theoretically\cite{kotliar88,general04}.
\begin{figure}[!!t]
\begin{center}
\includegraphics[width=9cm,height=4.0cm,angle=-0]{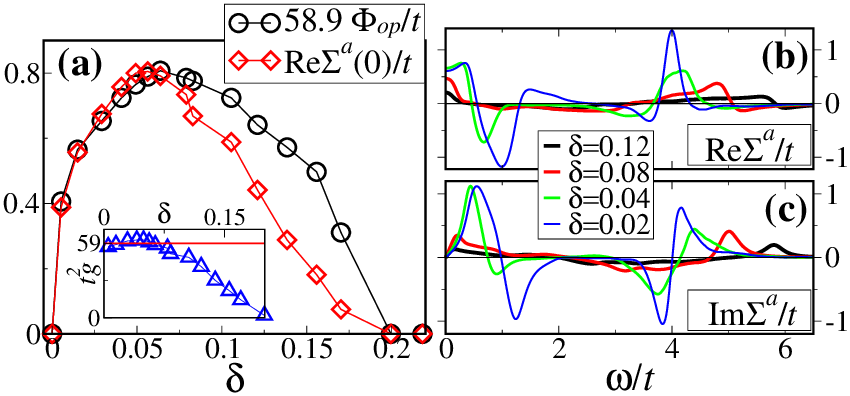}
\caption{(color online). The superconducting order parameter
$\Phi_{op}$ (multiplied by $t^2 g= 58.9$ shown in the inset) and
the anomalous self-energy value Re$\Sigma^{a}(\omega=0)$ are
displayed as a function of doping $\delta$ (a). The
Re$\Sigma^{a}(\omega)$ (b) and Im$\Sigma^{a}(\omega)$ (c) are
displayed on a wide energy range $0< \omega< U/2$, from the OD
region ($\delta=0.12, 0.08$) to the UD region ($\delta=0.04,
0.02$).} \label{OP-selfano}
\end{center}
\end{figure}

In Fig. \ref{OP-selfano}a the superconducting order parameter $
\Phi_{op}\equiv \int_{-\infty}^{\infty} F_{12}(\omega) d\omega $
is displayed as a function of doping $\delta= 1-\langle n_{i}
\rangle$. $\Phi_{op}$ has the expected dome-like shape, with a
maximum around $\delta_{opt}\sim 0.07$; this locates the optimal
doping in our Lanczos-CDMFT calculation (a similar $\delta_{opt}$
is also obtained in the QMC-CDMFT results of Ref.\cite{haule07}).
This value is smaller than the $\delta_{opt}\sim 0.15$ observed
in cuprate materials, which require more complete models to be
described in detail. However, as with the cuprates, we shall
define the region of $\delta> \delta_{opt}$[$\delta<\delta_{opt}$]
as the OD region [UD region]. If we compare $\Phi_{op}$ with the
anomalous self-energy value $\Sigma^{a}(\omega=0)$ we can already
observe the unusual non-BCS behavior of the Cooper pairing in the
OD region. If BCS theory is a good approximation,
$\Sigma^{a}(0)\sim g \, \Phi_{op}$, where $g$ is the strength of
the pairing interaction\cite{fetter03}. But $\Sigma^{a}(0)$
roughly scales like $\Phi_{op}$ only in the UD region (with
$g\sim 58.9$), with no scaling possible in the OD region (as is
shown by the $\delta$-dependence of $g$ in the inset).

To clarify the observation above we investigated the $\omega$-dependence
of $\Sigma^{a}(\omega)$ (Fig. \ref{OP-selfano}b,c). As with previous
results\cite{haule07,maier08}, $\Sigma^{a}(\omega)$
shows features which extend to high energies ($\omega\sim 6t $).
Re$\Sigma^{a}$ assumes negative values; these are particularly
evident in the UD region (around $0.5 t< \omega< 3.5 t$),
indicating a repulsive pair interaction. In absolute value $\Sigma^{a}$
decreases with doping, although Fig. \ref{OP-selfano}a shows that
Re$\Sigma^{a}(\omega\to 0)$ increases in the UD region
(in agreement with the QMC-CDMFT results\cite{haule07}).
\begin{figure}[!!t]
\begin{center}
\includegraphics[width=9cm,height=7.0cm,angle=-0]{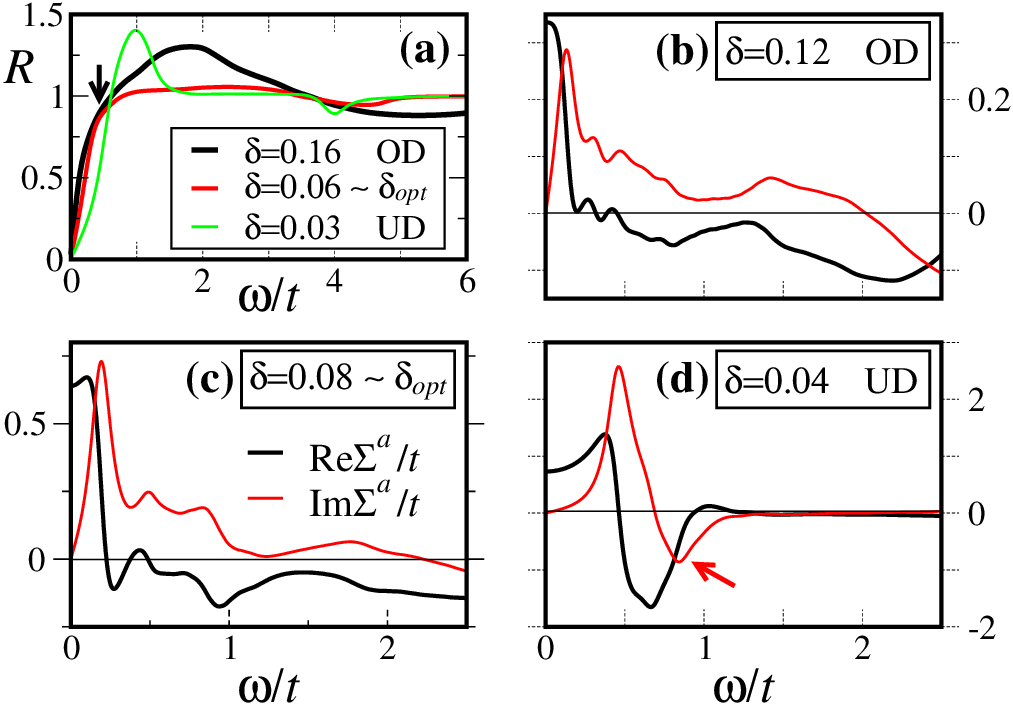}
\caption{(color online). The ratio $R$ (defined in formula
\ref{Rw}) vs. $\omega$ (a). The low-energy close-up ($\omega<
W^{\ast}$) of Re$\Sigma^{a}(\omega)$ and Im$\Sigma^{a}(\omega)$
is displayed in the OD region (b), close to optimal doping
$\delta_{opt}$ (c) and in the UD region (d).} \label{selfano}
\end{center}
\end{figure}

We now look in detail at the evolution of $\Sigma^{a}$ across the
phase diagram. We first employ Cauchy-Kramers-Kroenig relations
and define the ratio\cite{poilblanc05,maier08}
\begin{equation}
R(\omega)= \frac{1}{\pi}\frac{1}{\hbox{Re}\Sigma^{a}(0)}
\int_{0}^{\omega} \, d\nu \, \frac{\hbox{Im}\Sigma^{a}(\nu)}{\nu}
\label{Rw}
\end{equation}
$R(\omega)$ measures the contribution to the low-energy
superconducting pairing Re$\Sigma^{a}(\omega=0)$ obtained from the
range of frequency up to $\omega$. We identify three distinct
doping-dependent regimes (Fig. \ref{selfano}a). (i) In the OD
region, $R(\omega)$ monotonically increases to $\omega\sim
W^{\ast}\sim 2t$, and then decreases, showing a wide hump which
overshoots one, centered around $\omega\sim 1.5 t$. The small
hump also visible (small arrow) around $\omega\sim 0.5 t$
indicates the presence of a boson-like contribution to
$\Sigma^{a}$. (ii) Close to optimal doping $\delta_{opt}$, the
range of increasing monotonicity of $R(\omega)$ is reduced to
$\omega\leq t$. For $\omega> t$, $R(\omega)$ stays roughly flat
around one. (iii) In the UD region, a narrow peak overshooting
one appears at low frequency $\omega\leq t$. This form of
$R(\omega)$ is similar to that expected from a boson-mediated
pairing mechanism\cite{scalapino69,maier08}.

$R(\omega)$ shows that the pairing builds up for $\omega\leq
W^{\ast}$ and hints at the presence of a low-energy boson.
Bosonic modes have actually been detected in high-Tc
materials\cite{yazdani08,heumen08,maksimov08}.
Figs. \ref{selfano}b,c,d, which show the low energy part
of $\Sigma^{a}$, therefore merit close
examination. In ME theory, the boson exchange between pairing
electrons resonates as $\omega\to \omega_o^-$,
a characteristic frequency. This fact is made evident by a hump
in Re$\Sigma^{a}(\omega)$. For $\omega \geq \omega_o$ however,
the probability of emission of a real boson increases; this is
marked by Re$\Sigma^{a}(\omega)$ changing sign and by
Im$\Sigma^{a}(\omega)$ acquiring a peak. This peak occurs at
$\Omega\sim \omega_o+ \hbox{Re} \Delta_{tot}(0)$, $\hbox{Re}
\Delta_{tot}(0)$ being the lowest particle-excitation
available in the superconductor (the superconducting
gap in a standard BCS superconductor). In the OD region
(Fig. \ref{selfano}b), $\Sigma^{a}$ has the ME form
described above at low $\omega$, but it also has other features
at higher frequencies. The pair interaction is repulsive for
$\omega\geq 0.5 t$, where Re$\Sigma^{a}$ changes sign.
The Im$\Sigma^{a}$ however has a long positive tail until
$\omega\sim W^{\ast}\sim 2 t$, which means that the pairing mechanism
involves states in the renormalized band of width $W^{\ast}$.
The superconducting pairing in the OD region cannot therefore be
described simply within the ME theory, which typically shows only
one or few characteristic boson-frequencies (where Im$\Sigma^{a}$
shows well defined peaks). Around $\delta_{opt}$ (Fig.
\ref{selfano}c) the long tail in Im$\Sigma^{a}$
reduces strongly. 
In the UD region (Fig. \ref{selfano}d), the tail disappears, and
$\Sigma_{ano}$ assumes an ME shape (in line with the findings of
Fig. 1), marking a change in the pairing mechanism.
\begin{figure}[!!t]
\begin{center}
\includegraphics[width=9cm,height=7.0cm,angle=-0]{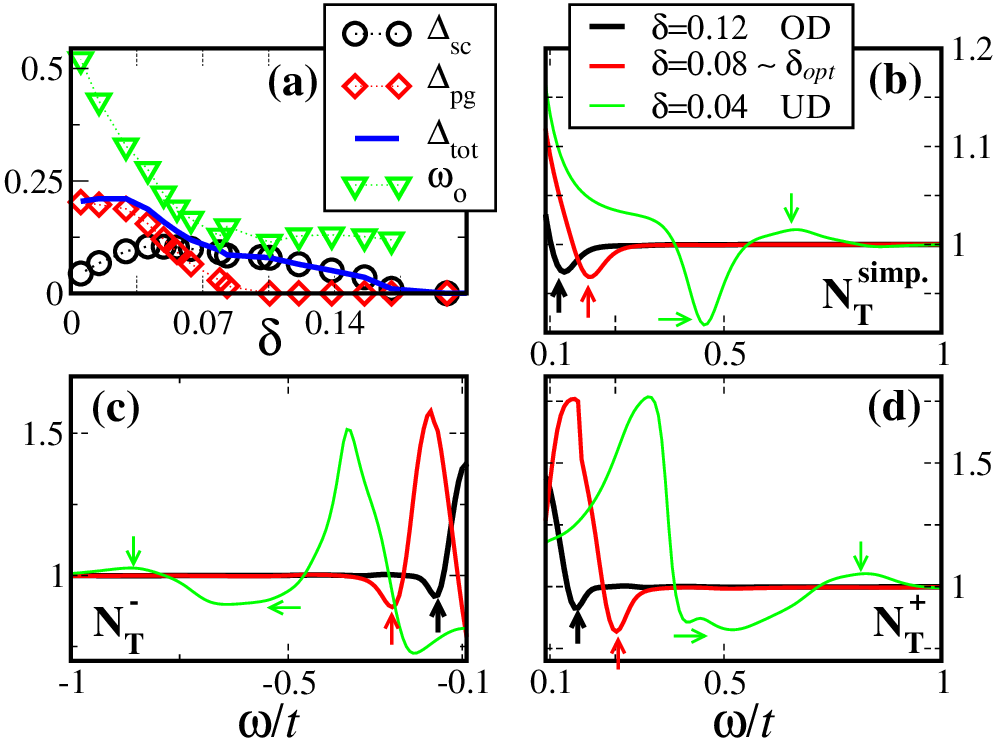}
\caption{(color online). The superconducting gap $\Delta_{sc}$,
the pseudo-gap $\Delta_{pg}$,
$\Delta_{tot}=\sqrt{\Delta_{sc}^{2}+\Delta_{pg}^{2}}$ and the
characteristic boson-frequency $\omega_o$ are displayed as
a function of doping $\delta$ (a). The conductance ratio
$N_T^{simp.}$ (b) is a simple ideal case (Eq. \ref{NT}), where
only the $\Sigma^{a}$-dependence is relevant. $N_T^{-}$ (c) and
$N_T^{+}$ (d) represent more realistic-material case, obtained
via a periodization\cite{marce08b} in momentum space of the CDMFT
results.} \label{wo-pg}
\end{center}
\end{figure}

We now link the evolution of the pairing function with the
relevant energy-scales of the system. In BCS theory there is only
one such scale, the superconducting gap $\Delta_{sc}\sim \omega_o
\hbox{e}^{-1/(N_o g)}$ (where $N_o$ is the density of states at
the Fermi level), proportional to the characteristic
boson-frequency $\omega_o$. In our results, however, there are
two relevant energy-scales\cite{marce08,marce08b}. The first
derives from Re$\Sigma^{a}$ and represents the superconducting gap
$\Delta_{sc}\sim Z_{k_{nod}}\, \hbox{Re}\Sigma^{a}(0)$ (where
$Z_{k_{nod}}$ is the quasiparticle
residue\cite{marce08,marce08b}). In our theory, $\Delta_{sc}$ can
be measured in the regions of momentum space close to ``nodal''
points $\mathbf{k}_{nod} \sim
(\pm\frac{\pi}{2},\pm\frac{\pi}{2})$ by
photo-emission\cite{letacon06,tanaka06}; its behavior is
non-monotonic with doping (Fig. \ref{wo-pg}a). The second energy
scale $\Delta_{pg}$ arises in the UD region from the normal
component $\Sigma^{nor}$ (see
Ref.\cite{stanescu06,marce08,marce08b}); it is connected with the
pseudo-gap observed in the normal state, and it is monotonic with
$\delta$. It is convenient to define a total gap\cite{marce08b}
$\Delta_{tot}= (\Delta_{pg}^{2}+ \Delta_{sc}^{2})^{\frac{1}{2}}$,
which is measured in photoemission\cite{letacon06,tanaka06} close
to the ``antinodal'' points $\mathbf{k}_{anod} \sim (0,\pm\pi)$
or $(\pm \pi,0)$. In the OD region $\Delta_{sc}\equiv
\Delta_{tot}$, but in the UD region the contribution of
$\Delta_{pg}$ at the antinodes is most
important\cite{marce08,marce08b}. To clarify whether the
boson-feature observed in Fig. \ref{selfano} can be related to
any of these gaps, we estimate its characteristic frequency
$\omega_o\sim \Omega-\Delta_{tot}(k_{anod})$ (at $\omega=\Omega$
we locate the first maximum of Im$\Sigma^{a}(\omega)$ in Fig.
\ref{selfano}b,c,d) and plot it as a function of doping $\delta$.
We find that $\omega_o$ follows the behavior of the pseudo-gap
$\Delta_{pg}$ (suggestive of the experimental results in
Ref.\cite{yazdani08}) rather than the superconducting gap
$\Delta_{sc}$. Coming from the OD region, $\omega_o$ is roughly
constant, just as $\Delta_{pg}=0$, while $\Delta_{sc}$ increases.
In the UD region, $\omega_o$ monotonically increases as doping is
reduced; it follows the rise of $\Delta_{pg}$ rather than the
fall of $\Delta_{sc}$, in spite of the ME form of $\Sigma^{a}$.
This is due to the fact that, in the UD region, negative values
appear in Im$\Sigma^{a}$ (oblique arrow in Fig. \ref{selfano}d)
which give a negative contribution to the Re$\Sigma^{a}(0)$ (Eq.
\ref{Rw}) and break the Cooper pairs.

In our interpretation\cite{stanescu06,marce08,marce08b},
these findings are Mott-related phenomena. At optimal doping
$\delta_{opt}$ the antinodal regions of momentum space undergo a
Mott transition ($\Delta_{pg}\geq 0$), while the nodal regions
retain a metallic character. These
ideas\cite{haule07,ferrero-2008} are reminiscent of the orbitally
selective Mott transition in multi-band models. Close to the Mott
transition, antiferromagnetic fluctuations are
important\cite{maier08,woo06,brehm08}, and can mediate a pairing
channel $\langle c_{i\uparrow} c_{j\downarrow} \rangle$ to form
Cooper pairs\cite{maier08}. The proximity to the ``antinodal''
Mott transition results in a pairing process involving electrons
on a wide energy scale (with Im$\Sigma^{a}$ positive up to
$\omega\sim W^{\ast}$). The high-Tc mechanism therefore originates
as one approaches a Mott transition\cite{anderson87}. Once this
latter has taken place at $\delta_{opt}$, a normal-component
particle-hole $\langle c^{\dagger}_{i\sigma} c_{j\sigma} \rangle$
channel is opened, destroying the Cooper pairing at
higher energies ($\Delta_{sc}$ decreases) and favoring the
rise of $\Delta_{pg}$. This is marked by the appearance of
non-BCS properties in the normal component together with
a pairing function of a more standard ME form.

Finally, let us look at how the doping-dependent description of the
pairing function can be measured in experiments. In ME
superconductivity it was possible to relate the behavior of
the superconducting pairing to STM measurements\cite{scalapino69} of
the ratio $N_T$, between the superconducting and normal-state
tunneling conductances $dI(\omega)/dV$, obtained under the same
conditions:
\begin{eqnarray}
&N_T(\omega)& = \frac{dI(\omega)/dV
\left|_{sc}\right.}{dI(\omega)/dV \left|_{nor} \right.}=
\frac{N_{sc}(\omega)}{N_{nor}(\omega)}\approx \label{NT}\\
&N_T^{simp.}(\omega)& = 1+\frac{Z_{k_{nod}}}{4 \omega^{2}}
\left[\hbox{Re}\Sigma^{a}(\omega)^2-
\hbox{Im}\Sigma^{a}(\omega)^{2}\right] \nonumber
\end{eqnarray}
where $N_{sc[nor]}(\omega)= \frac{1}{\pi} \sum_k
\hbox{Im}G(k,\omega)$ is the superconducting [normal-state] local
density of states. $N_T^{simp.}$ is the ideal $\omega\to\infty$
limit, obtained in the simplest ME analysis of a d-wave
superconductor\cite{scalapino69,yazdani08} neglecting the momentum
dependence of the band-structure and taking
$N_{nor}(\omega)\approx N_{nor}(0)$ (a good approximation in BCS
superconductors). $N_T^{simp.}$ has the advantage of being an
explicit function of the superconducting gap
$\Delta_{sc}(\omega)\sim Z_{k_{nod}} \Sigma^{a}(\omega)$. Its
behavior is shown in Fig. \ref{wo-pg}b. The sharp drop below
unity (due to the Re$\Sigma^{a}\to 0$ and Im$\Sigma^{a}$ acquiring
a maximum, see Fig. \ref{selfano}) marks the presence of the boson
(at $\omega= \Omega$, vertical-up and horizontal arrows in Fig.
\ref{wo-pg}b), both in the OD region (in agreement with
experiments\cite{yazdani08}) and in the UD region. It is not
possible to see the long tail of the Im$\Sigma^{a}$ in the OD
region, as the real and imaginary parts of $\Sigma^{a}$ cancel
out. We can however observe two clear features: (i) the sudden
increase of $\Omega$ in passing from the OD region to the UD
region, mainly due to the increase of $\omega_o$, as portrayed in
Fig. \ref{wo-pg}a, and (ii) the rise above unity at higher
frequencies (marked by a vertical-down arrow) in the UD region;
this is due to the pair-breaking effect of Im$\Sigma^{a}$ assuming
negative values (in a way opposite to the boson-drop, now
Im$\Sigma^{a}\to 0$, and Re$\Sigma^{a}$ is maximal). The
approximations used to extract the simple behavior of
$N_T^{simp.}$ no longer hold in the real-case situation, where
the momentum dependence of the band cannot be neglected and,
above all, $N_{nor}(\omega)$ is strongly $\omega$-dependent. We
performed a full momentum-energy-dependent calculation, however,
adopting a mixed-periodization scheme (introduced in
Ref.\cite{marce08,marce08b}), which allows us to extract a good
approximation for $G(k,\omega)$ from the cluster results, while
separating the normal component $N_{nor}$ from the
superconducting $N_{sc}$. The resulting $N_T^{+}$ for $\omega>0$
and $N_T^{-}$ for $\omega<0$ are displayed in Figs.
\ref{wo-pg}c,d. The curves are now more irregular and asymmetric
with respect to $\omega=0$, but the features (i) and (ii) which
we have discussed for $N^{simp.}_{T}$ are still present, showing
that it should be possible to observe them in
experiments\cite{yazdani08}.

We can conclude that the experiments and theories showing
unconventional properties in the UD region and more conventional
properties in the OD region of the phase diagram of high-Tc
superconductors do not tell the entire story. We find an
unconventional form of superconducting pairing in the OD region,
which evolves into a more standard form in the UD region. This
crossover takes place close to the maximum of the order
parameter, where a pseudo-gap appears in the one-electron density
of states. We argue that these phenomena are a consequence of a
Mott-like transition taking place at the antinodal points of
momentum space.
\begin{acknowledgments}
We would like to thank G. Kotliar, B. Kyung, Th. Maier, I. Paul,
D. Poilblanc, A.-M. Tremblay, R.S. Whitney for the helpful
discussions and P. Bruno, A. Cano, C. Pepin, T. Ziman for their
comments. We also thank R. Corner for proofreading this article.
\end{acknowledgments}



\end{document}